\documentclass[aps,prl,twocolumn,showpacs,twoside,amsmath, amsfonts,superscriptaddress]{revtex4}
\begin{document}
\title{T-duality in supersymmetric theory of disordered quantum systems}
\author{A. Ossipov}
\email{aossipov@ictp.trieste.it}
\affiliation{The Abdus Salam International Centre for Theoretical
Physics, P.O.B. 586, 34100 Trieste, Italy}
\author{V. E. Kravtsov}
\affiliation{The Abdus Salam International Centre for Theoretical
Physics, P.O.B. 586, 34100 Trieste, Italy}
\affiliation{Landau Institute for
Theoretical Physics, 2 Kosygina st., 117940 Moscow, Russia}
\begin{abstract}
A new super-symmetric representation for quantum disordered
systems is derived. This representation is exact and is dual to
that of the nonlinear sigma-model. The new formalism is tested by
calculating the distribution of wave function amplitudes in the 1d
Anderson model. The deviation from the distribution found for a
thick wire is detected near the band center $E=0$.

\end{abstract}
\pacs{72.15.Rn, 72.70.+m, 72.20.Ht, 73.23.-b}
\keywords{localization, supersymmetry, mesoscopic fluctuations}
\maketitle {\it Introduction.}--- The rigorous theory of quantum
disordered systems is based almost entirely on the
field-theoretical formulation of the nonlinear sigma-model (NL
$\sigma$M) \cite{Wegner79, Efet83}. The most successful variant of
this theory, the so called {\it supersymmetric} (SUSY) sigma-model
\cite{Efet83, Efet-book}, allows not only for a perturbative
treatment of weak localization and mesoscopic phenomena in the
delocalized (metallic) phase which was previously done by the
impurity diagrammatic technique \cite{GLKh} but also
applies to systems in the localized (insulator) phase provided that
the localization radius is sufficiently large. A celebrated
example is the problem of localization in a thick disordered wire
\cite{Efet-book, Mirl-rev} which is equivalent \cite{Mirl-Fyod91}
to the Gaussian ensemble of random banded matrices with a large
bandwidth $b\gg 1$.

However, despite of the great success of the nonlinear SUSY sigma-model
over the last two decades \cite{Efet-book} this approach is
limited. The limitation comes from the saddle-point approximation
that locks the norm of the supermatrix field $Q$ and thus
leads to a geometric constraint $Q^{2}=1$. It is precisely
the validity of this saddle-point approximation that requires a
quantum system to be in the metallic phase (at least for a
sufficiently small length scale), or the banded random matrices to
have a large bandwidth.

The best known example for which the NL$\sigma$M is insufficient
is the strictly one-dimensional disordered chain described (in the
position representation) by the {\it three-diagonal} matrix
Hamiltonian $H_{ij}=J_{i} \,\delta_{j,i\pm 1}+
\varepsilon_{i}\,\delta_{ij}$. The case $J_{i}=1$, and
$\varepsilon_{i}$ being a Gaussian random variable with zero mean
and the variance $w=\langle \varepsilon_{i}^{2}\rangle$ is known
as the 1d Anderson model. Its continuous counterpart for a weak
disorder $w\ll 1$ was studied in detail in the 70-th
\cite{Berez74, AbrRyzh, Mel} with the classic results on the
frequency dependent conductivity \cite{Mott, Berez74} and dc
conductance distribution \cite{AbrRyzh, Mel}. The special case
where $\varepsilon_{i}=0$ and $J_{i}$ is a Gaussian random
variable has been studied in the pioneer work by Dyson
\cite{Dyson} and attracted a lot of interest in recent years in
connection with the new symmetry classes \cite{AltZirn}. Yet
another broad class of systems which cannot be studied by the
NL$\sigma$M are systems described by the  {\it almost diagonal}
random matrix Hamiltonian \cite{YevKra} with the variance of the
diagonal entries $\langle H_{ii}^{2} \rangle$ much larger than
that for the off-diagonal ones $\langle |H_{i\neq j}|^{2}
\rangle$. Some of such systems with $\langle |H_{i\neq j}|^{2}
\rangle\sim |i-j|^{-2}$  attracted recently much attention in
connection with the multifractality of the critical eigenstates
\cite{MirFyodPL, KraMut, YevKra}.

The main goal of this work is to fill in the gap in our technical
ability to apply the SUSY method to quantum disordered systems.
Below we present a field-theoretical description that is free of
of the saddle-point approximation and thus applies to all the
above cases which are not amenable to the conventional
NL$\sigma$M. This description keeps all the Grassmann variables
and thus is different from another suggestion, the {\it
super-bosonization}, which was recently proposed in Ref.
\cite{SupBos} as an exact method instead of the NL$\sigma$M but is
suffering from uncertainties and incomplete definition of the target
space. The method we are proposing is similar in spirit to
that of Ref.\cite{SupBos} but has an advantage to be as close as
possible to the standard SUSY NL$\sigma$M. In particular, the new
supermatrix field $\hat{Q}$ can be diagonalized by a pseudo-unitary
transformation with separated Grassmann and commuting variables
\cite{Efet-book} similar to the one used in the NL$\sigma$M. This
allows to drastically simplify calculations with the final result
depending on few commuting variables only, as for the conventional
NL$\sigma$M.

The main property of the new description is that it is ${\it
dual}$ to the old one: the coupling matrix of the new theory is
just the inverse of the coupling matrix of the NL$\sigma$M.
However, the space of the matrices $\hat{Q}$ of the new theory (the $\hat{Q}$
target space) is different from the $Q$ target space of the
NL$\sigma$M. So, we are dealing not with the conventional duality
but with a kind of $T$-{\it duality} \cite{string} remotely
resembling the situation in the string theory.

As an illustration of the new formalism at work we consider the
problem of the eigenfunction statistics for a one-dimensional Anderson
model. We derive the transfer matrix integral recursive equation
following the same rout as the one used in the framework of the
SUSY nonlinear sigma-model for quasi-1d disordered wires
\cite{Efet-book} or banded random matrices \cite{Mirl-Fyod91}. For
weak disorder this integral equation reduces to the the
differential equation of exactly the same form as for a thick
disordered wire. This fact (noticed earlier in \cite{Efet-book,Mirl-rev})
implies that despite the absence of the diffusive regime in a
strictly 1d chain, the smooth envelope of a wavefunction at scales
much larger than the wavelength and the wire cross-section has
exactly the same statistics in a thick disordered wire and in a
strictly 1d disordered chain.

{\it The dual representation.}---For concreteness we consider the
matrix Hamiltonian of the unitary class ${\bf H}$ which entries
$H_{ij}$ are random independent Gaussian variables characterizing
by the variance matrix ${\bf g}_{ij}=\langle |H_{ij}|^{2}\rangle$
and the mean values $\langle H_{ij} \rangle=H^{(0)}_{ij}$. The
standard SUSY field-theoretical description \cite{Efet-book} aimed
to computing the retarded/advanced (R/A) Green's functions
$[E_{\pm}-{\bf H}]^{-1}$ begins with introducing the $\Psi$
functional
$S=\sum_{ij}\bar{\Psi}_{i}\,(E_{\pm}\,\delta_{ij}-H_{ij})\,\Psi_{j}$,
where $E_{\pm}=E\pm \tilde{\omega}/2$ and
$\tilde{\omega}=\omega+2i\eta$ with $\eta\rightarrow +0$. The {\it
ket }-supervector $\Psi=(S_{R},S_{A},\mu_{R},\mu_{A})^{\rm{T}}$ contains
both commuting complex variables $S_{R/A}$ and Grassmann
anti-commuting variables $\mu_{R/A}$. For convergence of the
Gaussian functional integrals containing $e^{iS[\Psi]}$ the {\it
bra }-supervector $\bar{\Psi}$ is given by
$\bar{\Psi}=\Psi^{\dagger}\Lambda$ where $\Lambda={\rm diag}\{1,-1,1,-1
\}$. The SUSY trick allows to do averaging over disorder prior to
doing the functional integral over $\Psi$. Thus we obtain $\langle
e^{iS[\Psi]}\rangle=e^{-F}$, where
\begin{eqnarray}
\label{F} F&=& \frac{1}{2}\sum_{i,j}{\bf g}_{ij}\,
{\rm Str}[\hat{Q}_{i}\hat{Q}_{j}]-\frac{i\tilde{\omega}}{2}\sum_{i}{\rm Str}[\Lambda
\hat{Q}_{i}]\\ \nonumber &-&iE\sum_{i}{\rm Str}[\hat{Q}_{i}] + F_{kin}.
\end{eqnarray}
Here $F_{kin}=i\sum_{ij}\bar{\Psi}_{i}\,H^{(0)}_{ij}\,\Psi_{j}$
and
\begin{equation}
\label{Q}
\hat{Q}=\Psi\otimes\,\bar{\Psi}=\Psi\otimes\,\Psi^{\dagger}\,\Lambda.
\end{equation}
So far we did not deviate from the standard rout except for
introducing the new notations Eq.(\ref{Q}). However, the
integration measure is still unchanged $d\mu[\Psi]
=dS_{R}dS_{A}dS^{*}_{R}dS^{*}_{A}d\mu_{R}d\mu_{A}d\mu^{*}_{R}d\mu^{*}_{A}$.

Now we do the crucial step. Instead of performing the
Hubbard-Stratonovich transformation which decouples  the quartic
in $\Psi$ term in Eq.(\ref{F}) by the supermatrix $Q$ and applying
the subsequent saddle point approximation which leads to the
sigma-model constraint $Q^{2}=1$ we change the variables so that
in the new variables the supermatrix $\hat{Q}$ is parameterized
similar to the supermatrix $Q$ of the sigma-model:
\begin{eqnarray}
\label{ChoV} S_{R/A}&=&\pm i\sqrt{\lambda_{1/2}}\,\,e^{\pm
i\varphi/2+i\Omega}\,(1-\frac{1}{2}\chi^{*}_{R/A}\chi_{R/A})\\ \nonumber
\mu_{R/A}&=& \pm i \sqrt{\lambda_{1/2}}\,\,e^{\pm
i\varphi/2+i\Omega}\,\chi_{R/A},
\end{eqnarray}
where $\lambda_{1/2}\geq 0$, $0\leq \varphi\leq 2\pi$,
$0\leq\Omega\leq \pi$ and $\chi_{R/A}, \chi^{*}_{R/A}$ are the new
Grassmann variables.

One can check that in new variables the supermatrix $\hat{Q}$
defined in Eq.(\ref{Q}) can be represented in the form where
Grassmann and commuting variables are separated by factorization:
\begin{equation}
\label{param} \hat{Q}=\left(\begin{matrix}u_{R} & 0 \cr 0 &
u_{A}\cr
\end{matrix} \right)\;\left(\begin{matrix}\hat{D}_{RR} & \hat{D}_{RA} \cr
 \hat{D}_{AR} &
\hat{D}_{AA}\cr
\end{matrix}\right)\,\left(\begin{matrix}u^{-1}_{R} & 0 \cr 0 &
u^{-1}_{A}\cr \end{matrix}\right).
\end{equation}
The factorization Eq.(\ref{param}) is also one of the basic
properties of the Efetov's parameterization in the sigma-model
\cite{Efet-book}. Moreover, the corresponding matrices $u_{R/A}$
are essentially the same:
\begin{equation}
\label{U} u_{R/A}=\left(\begin{matrix}1-\frac{1}{2}\chi_{R/A}^{*}\chi_{R/A}
& -\chi_{R/A}^{*} \cr \cr \chi_{R/A} &
1+\frac{1}{2}\chi_{R/A}^{*}\chi_{R/A}\cr
\end{matrix}\right)_{BF}.
\end{equation}
The difference is in the form of the matrix $\hat{D}$. Like in the
case of Efetov's parameterization it is diagonal in the boson-fermion
(BF) space.
However, in contrast to the sigma-model {\it only} the bosonic
sector is nonzero:
\begin{equation}
\label{Db} \hat{D}_{BB}=\left(\begin{matrix}\lambda_{1}&
\sqrt{\lambda_{1}\lambda_{2}}\,e^{i\varphi} \cr
-\sqrt{\lambda_{1}\lambda_{2}}\,e^{-i\varphi} & -\lambda_{2}\cr
\end{matrix}\right)_{RA},\;\;\;\;\hat{D}_{FF}=0.
\end{equation}
One can see that the matrix $\hat{D}_{BB}$ can be diagonalized by
the same {\it pseudo-unitary} transformation as the matrix
$D_{BB}$ of the sigma-model:
$\hat{D}_{BB}=\hat{T}\,\hat{B}_{0}\,\hat{T}^{-1}$, where $\hat{T}\in
U(1,1)$. However, the diagonal matrix
$\hat{B}_{0}=|\lambda_{1}-\lambda_{2}|\,{\rm diag}
(\theta(\lambda_{1}-\lambda_{2}),\theta(\lambda_{2}-\lambda_{1}))_{RA}$
is different from the corresponding diagonal matrix $\Lambda_{BB}$
of the sigma-model and contains a new degree of freedom,
the difference $\lambda_{1}-\lambda_{2}\in \mathbb{R}$. This means
that the full symmetry of the target space is ${\bf
\frac{U(1,1)}{U(1)\times U(1)}\times \mathbb{R}}$ rather than ${\bf
\frac{U(1,1)}{U(1)\times U(1)}\times \frac{U(2)}{U(1)\times U(1)}}$
as for the sigma-model.

To complete the procedure of changing variables from $\Psi$ to
$\hat{Q}$ we have to compute the Jacobian of the transformation
Eq.(\ref{ChoV}) and rewrite $F_{kin}$ in Eq.(\ref{F})
in
terms of $\hat{Q}$. The first task is straightforward and the
result is:
\begin{equation}
\label{Jac} d\mu\equiv d\hat{Q}d\Omega=
d\chi_{R}d\chi_{R}^{*}d\chi_{A}d\chi_{A}^{*}d\varphi d\lambda_{1}
d\lambda_{2}\;\frac{d\Omega}{\lambda_{1}\lambda_{2}}.
\end{equation}
To accomplish the second one we note that
$\bar{\Psi}_{i}=\bar{\varphi_{i}}\,e^{-i\Omega_{i}}$ and
$\Psi_{j}=\varphi_{j}\,e^{i\Omega_{j}}$ where
$\varphi_{i}=\Psi_{i}|_{\Omega_{i}=0}$. Then each link $\{ij \}$
enters in the last term of Eq.(\ref{F}) as
$iV_{ij}=i(p_{ij}\,e^{-i\Omega_{ij}}+p_{ji}\,e^{i\Omega_{ij}})$,
where $p_{ij}=\bar{\varphi}_{i}\varphi_{j}\,H^{(0)}_{ij}$, and
$\Omega_{ij}=\Omega_{i}-\Omega_{j}$. Then assuming that the
observable of interest is gauge-invariant and thus independent of
$\Omega$ and integrating $e^{-i\sum_{ij}V_{ij}}$ over all
$\Omega_{ij}\in [-\pi,\pi]$ one obtains for the last term in
Eq.(\ref{F}):
\begin{equation}
\label{Kin}
F_{kin}=-\sum_{j>i}\ln\left\{2\pi\,J_{0}\left(2\sqrt{|H^{(0)}_{ij}|^{2}\,
{\rm Str}[\hat{Q}_{i}\hat{Q}_{j}]} \right)\right\}.
\end{equation}
Note, however, that in order to arrive at Eq.(\ref{Kin})
integration over {\it all} $\Omega_{ij}$ should be independent.
This is the case when the links provided by $H^{(0)}_{ij}$ form a
lattice with {\it no loops}. The simplest example considered below
is a one-dimensional lattice with only nearest neighbor
connections $H^{(0)}_{i,i\pm 1}$.

 The new supermatrix field theory given by
Eqs.(\ref{F}),(\ref{Kin}) together with the parameterization of the
supermatrix $\hat{Q}$, Eqs.(\ref{param})-(\ref{Db}) is the main
result of the paper. It has to be compared with the standard
supermatrix sigma-model \cite{Efet-book}. In the case $\langle
H_{ij}\rangle =0$ and ${\bf g}_{ij}={\bf g}(\mid i-j\mid)$ the
 action reads \cite{Mirl-Fyod91}:
\begin{equation}
\label{sigma-mod} F_{sm}=-\frac{(\pi \nu\,A_{0})^{2}}{2}\sum_{ij}{\bf
g}^{-1}_{ij}\,{\rm Str}[Q_{i}Q_{j}]-\frac{i\pi\nu\,\tilde{\omega}}{2}\sum_{i}{\rm Str}[\Lambda
Q_{i}],
\end{equation}
where $\nu(E)=\frac{1}{4\pi A_{0}}\sqrt{8A_{0}-E^{2}}$ is the mean
density of states and $A_{0}=\sum_{i}{\bf g}_{ij}$. Comparing
Eq.(\ref{sigma-mod}) with Eq.(\ref{F}) one notices that the
coupling matrix ${\bf g}$ in Eq.(\ref{F}) is replaced in
Eq.(\ref{sigma-mod}) by the {\it inverse} coupling matrix ${\bf
g}^{-1}$. That is why the new representation is {\it dual} to the
standard sigma-model representation. However the target spaces of
the supermatrices $\hat{Q}$ and $Q$ in Eq.(\ref{F}) and
Eq.(\ref{sigma-mod}) are different. The $Q$-field is constrained:
$Q^{2}=1$ and ${\rm Str}\, Q=0$. That is why the term $E\,{\rm Str}\,Q$ cannot
arise and the energy $E$ enters as an inessential parameter. There
is no such constrains for the supermatrix $\hat{Q}$ in
Eq.(\ref{F}) and the terms $E\,{\rm Str}\,\hat{Q}$ and ${\bf
g}_{ii}\,{\rm Str}\,\hat{Q}_{i}^{2}$ are important. In particular this leads
to peculiar properties of the 1d disordered chain at the point
$E=0$ \cite{Titov, Alt-Lys}, especially in the {\it chiral} case
\cite{Dyson} where diagonal matrix elements (on-site energies)
$H_{ii}$ do not fluctuate.

{\it 1d Anderson model.}---We illustrate the new method by
considering the statistics of an eigenfunction $\psi_{n}(i)$
corresponding to the eigenvalue $E_{n}$ for the 1d Anderson model.
This problem has been studied earlier \cite{Kolokolov} by a
different method which was not based on the SUSY trick. The 1d
Anderson model corresponds to the field theory
Eqs.(\ref{F}),(\ref{Kin}) where the coupling matrix ${\bf
g}_{ij}=w\,\delta_{ij}$ is diagonal and
$H^{(0)}_{ij}=\delta_{i,j+1}+\delta_{i,j-1}$. To compute the
distribution function $P_{i}(t)=\nu^{-1}(E)\,\langle\sum_{n}
\delta(t-|\psi_{n}(i)|^{2})\,\delta(E_{n}-E) \rangle$ we follow
the procedure used in \cite{Efet-book, Mirl-rev}  for thick
quasi-1d wires where the sigma-model applies. It starts by the
exact expression for the $q$-th moment of this distribution
\begin{equation}
\label{I-Y}
I_{q}(i)=\frac{1}{4\pi\nu(E)}\,\frac{1}{(q-2)!}\int_{0}^{\infty}
ds\,s^{q-2}\int_{-\infty}^{+\infty}dv\;Y_{i}(s,v).
\end{equation}
in terms of the function $Y(\hat{Q}_{i})=\int \prod_{\ell \neq
 i}d\hat{Q}_{\ell}\,e^{-F(\hat{Q})}$.
The functional $F$  is given by Eq.(\ref{F}), where
$\tilde{\omega}=2\eta\to+0$ and
$F_{kin}=\sum_{i}F_{kin}(\hat{Q}_{i},\hat{Q}_{i+1})=-\sum_{i}\ln\,2\pi
J_{0}(2\sqrt{{\rm Str}\,\hat{Q}_{i}\hat{Q}_{i+1}})$.

As usual \cite{Mirl-rev} in the 1d chain of the length $N$ the
function $Y_{i}=W_{i}W_{N-i}\,e^{-L(\hat{Q}_{i})}$ is found from
the recursive relation:
\begin{equation}
\label{rec} W_{j+1}(\hat{Q})=\int
d\mu(\hat{Q}')\,e^{-F_{kin}(\hat{Q},\hat{Q}')-L(\hat{Q}')}\,\,W_{j}(\hat{Q}'),
\end{equation}
where $L(\hat{Q})=\frac{w}{2}\,{\rm Str}\,\hat{Q}^{2}+ {\rm
Str}\,\Lambda\hat{Q}-iE\,{\rm Str}\,\hat{Q}$, and the boundary
condition $W_{j=1}(\hat{Q})=1$ is assumed.

One can show that the integral in Eq.(\ref{rec}) does not depend on the
Grassmann variables $\chi$ and on the angle $\varphi$. Thus $W_i$ and $Y_{i}$
are functions of two ($\lambda_1,\lambda_2$) of the seven variables only.
This property is based on the form of the parameterization
Eqs.(\ref{param})-(\ref{Db}) and similar to that known in the sigma-model
approach \cite{Zirnbauer,Efet-book,Mirl-rev}, where the corresponding
functions depend on the non-compact $\lambda_B$ and the compact
$\lambda_{F}$ angles only. In the limit $\eta\to +0$,  $W_i$ remains a
function of {\it two} new  variables $v=\lambda_1-\lambda_2$ and
$s=\eta (\lambda_1+\lambda_2)$. At this point the difference in the
symmetries of the target spaces in our approach and the sigma-model becomes
important. In the latter case $W_i$ depends on one rescaled
variable $(2\eta)\,\lambda_B$ only, while the dependence on $\lambda_F$
disappears \cite{Zirnbauer,Efet-book,Mirl-rev}. The fact that the recursive
relation Eq.(\ref{rec}) and the Fokker-Plank equation that follows
from it depend on {\it two} variables is generic to all methods
in the theory of 1d disordered chains \cite{Pastur} thus
demonstrating that the SUSY approach is the {\it minimal}
description. We will see that the dependence on $v$ reflects
statistics of eigenfunction oscillations at scales of the order of
the de Broglie wavelength $k_{F}^{-1}$ which are essentially
different in a thick wire and in a 1d chain.

After integration over the angle $\varphi'$ and over the Grassmann
variables according to a convention $\int
\chi'_{R/A}\,d\chi'_{R/A}=\frac{1}{\sqrt{2\pi}}$ and going to the
limit $\eta\rightarrow 0$ one obtains an exact recursive equation:
\begin{eqnarray}
\label{rec1}
W_{j+1}(s,v)&=&\frac{s^{\frac{1}{2}}}{2\pi}\,\int_{\infty}^{+\infty}dv'
\int_{0}^{\infty}\frac{ds'}{(s')^{\frac{3}{2}}}\,W_{j}(s',v') \\
\nonumber &\times&
\cos\left[\sqrt{ss'}\,\left(\frac{v'}{s'}+\frac{v}{s}\right)
\right]\, e^{-s'-\frac{w}{2}\,v'^{2}+iE v'}.
\end{eqnarray}
Similar equations have been derived earlier
\cite{MirlFyodNucl,AbouAndThoul} in study of the localization
transition on the Bethe lattice. Eq.(\ref{rec1}) is valid for any
strength of disorder $w$ and for an arbitrary energy $E$. However
in the limit of weak disorder $w\ll 1$ it can be drastically
simplified using the {\it scale separation} $k_{F}^{-1}\ll
\xi_{loc}$ where $\xi_{loc}$ is the localization length. To this
end we first consider the disorder-free case. Setting $w=0$ in
Eq.(\ref{rec1}) one notices that the Fourier-transform
$\tilde{W}_{j}(s',q')=\int dv'\, W_{j}(s',v')\,e^{iv'q'}$
naturally arises. Remarkably, the equation for
$\tilde{W}_{j}(s,q)$ turns out to be an {\it algebraic} and not an
integral equation:
$\tilde{W}_{j+1}(s,q)=\frac{1}{q^{2}}\,e^{-sq^{2}}\,\tilde{W_{j}}\left(sq^{2},E-
\frac{1}{q}\right)$.
This functional equation can be simplified if we introduce new
variables  $s=z\cos^{2}(\phi/2+k)$ and $q=\cos(\phi/2)/\cos(\phi/2
+k)$ and a new function:
\begin{equation}
\label{new}
\Phi_{j}(z,\phi)=\tilde{W}_{j}\left(s(z,\phi),q(z,\phi)\right)\,\frac{\sin
k}{2\cos^{2}(\phi/2+k)},
\end{equation}
where $k$  defined by the relation $E=2\cos k$ is a momentum for
plane waves in
 the tight-binding model.

In new variables we obtain:
\begin{equation}
\label{alg2}
\Phi_{j+1}(z,\phi)=e^{-z\,\cos^{2}(\phi/2)}\,\Phi_{j}(z,\phi-2k).
\end{equation}
One can see that the variable $\phi/2$ is related with the phase
of the wave function as it changes by $k$ in passing from a
lattice site $j$ to the neighboring one. For an irrational $k/\pi$
as one moves along the chain at a distance $\xi_{loc}\gg 1$, the
phase $\phi$ sweeps with the constant density through the entire
interval $[0,2\pi]$.
Therefore $\phi$ is the true "fast"
variable \cite{Pastur} in contrast to the "slow" variable $z$.
Thus one comes
 to an idea \cite{Pastur} of {\it averaging} over $\phi$ and replacing
$\Phi_{j}(z,\phi)$ in Eq.(\ref{new}) by its average (the "fast
phase" ansatz):
$\Phi_{j}(z)\equiv\overline{\Phi_{j}(z,\phi)}=\frac{1}{2\pi}\int_{0}^{2\pi}d\phi\;\Phi_{j}(z,\phi)$.

Now we take into account weak disorder by expanding
$e^{-\frac{w}{2}\,v'^{2}}=1-\frac{w}{2}\,v'^{2}=1+\frac{w}{2}\,\partial^{2}/\partial
q'^{2}$ in Eq.(\ref{rec1}). We also assume $z\sim w\ll 1$ and
expand $e^{-z\,\cos^{2}(\phi/2)}\approx 1-z\,\cos^{2}(\phi/2)$.
Then doing the Fourier transform over $v$ and $v'$, switching to
the new variables $z,\phi$ and averaging over $\phi$ one obtains
for $\Phi_{j+1}(z)-\Phi_{j}(z)=
c_{1}\,z\Phi_{j}(z)+c_{2}\,z^{2}\frac{\partial^{2}\Phi_{j}}{\partial
z^{2}}+c_{3}\,\left(\Phi_{j}-z\,\frac{\partial\Phi_{j}}{\partial
z}\right)$,
where $c_{1}=-\overline{\cos^{2}(\phi/2)}=-1/2$,
$c_{2}=\frac{w}{\sin^{2}k}\,\overline{(1-\cos^{2}\phi)}=\frac{w}{2\sin^{2}k}$,
and
$c_{3}=\frac{w}{\sin^{2}k}\,\overline{(1+\cos\phi)(1-2\cos\phi)}=0$.
Then switching to the continuous spacial variable
$\tau=j/\xi_{loc}$ with $\xi_{loc}=\frac{2\sin^{2}k}{w}$ and the
variable $x=z\xi_{loc}/2$ we obtain the following Fokker-Plank
equation subject to the boundary condition $\Phi_{\tau=0}(x)=1$:
\begin{equation}
\label{FP} \frac{\partial
\Phi_{\tau}(x)}{\partial\tau}=\left(x^{2}\frac{\partial^{2}}{\partial
x^{2}} -x\right)\,\Phi_{\tau}(x).
\end{equation}
The fact that $c_{3}=0$ depends crucially on the assumption on the
homogeneous distribution of the fast phase $\phi$ which is related
with the single-parameter scaling. As a matter of fact the
distribution of phases $F(\phi;E/w)$ depends on energy and is
homogeneous $F(\phi;\infty)=\frac{1}{2\pi}$ only for irrational
$k$ away from the band center $E=0$. At the band center
$F(\phi;0)\propto (3+\cos\phi)^{-1/2}$ is different
\cite{OssKotGeis} and we obtain $c_{1}= 1-2E/K\approx -0.457$,
$c_{2}\frac{\sin^{2}k}{w}=(-16/3+8E/K)\approx 0.494$ and
$c_{3}\frac{\sin^{2}k}{w}=(-26/3+12E/K)\approx 0.075$, where
$E=E(1/2)$ and $K=K(1/2)$ are elliptic integrals. Thus at the band
center $|E|\ll w$ Eq.(\ref{FP}) should be modified by including
the combination $\Phi - x\,\partial\Phi/\partial x$ proportional
to $c_{3}$.

Finally we have to switch to new variables in Eq.(\ref{I-Y}) which
involves the combination
$\tilde{W}_{i}(s,q)\,\tilde{W}_{N-i}(s,E-q)$.
After replacing $\tilde{W}$ by $\Phi$ according to Eq.(\ref{new}),
using the fast phase ansatz $\Phi_{j}(z,\phi)\rightarrow
\Phi_{\tau}(x)$ and integrating over $\phi$ one obtains for
$|E|\gg w$:
\begin{equation}
\label{I-fin}
I_{q}(\tau)=\frac{(2q-1)!!}{(q!)^{2}}\,\int_{0}^{\infty}dy\,q(q-1)\,y^{q-2}\,
Y_{\tau}(\xi_{loc}y),
\end{equation}
where we used the relation $2\pi\nu(E)=dk/dE=1/(2\sin k)$ and the
definition
$Y_{\tau}(x)=\Phi_{\tau}(x)\,\Phi_{\frac{N}{\xi_{loc}}-\tau}(x)$.
The result given by Eqs.(\ref{FP}),(\ref{I-fin}) is in a perfect
agreement with the one obtained in Ref.\cite{Kolokolov} by a
completely different method.

Remarkably, the integral in Eq.(\ref{I-fin}) and the Fokker-Plank
equation Eq.(\ref{FP}) coincide completely \cite{Mirl-rev} with
the corresponding equations \cite{Efet-book, Mirl-rev} determining
statistics of wave functions for a {\it multi-channel} disordered
wire. The $q$-dependent factor in front of the integral in
Eq.(\ref{I-fin}) is due to the difference in statistics of
oscillations of almost ballistic (in a strictly 1d system) and
fully chaotic (in a thick wire) wave functions at a small scale of
the order of the de Broglie wavelength. This coincidence is a
manifestation of the single-parameter scaling and it breaks down
near the band center $|E|\ll w$.

\end{document}